\documentclass[prl,onecolumn,aps,superscriptaddress,longbibliography]{revtex4-1}
\usepackage{graphicx}
\usepackage[fleqn]{amsmath}
\usepackage{amssymb}
\usepackage[usenames]{color}
\usepackage[normalem]{ulem}
\definecolor{goodgreen}{rgb}{0.1,0.5,0}
\definecolor{goodred}{rgb}{0.7,0,0}
\usepackage{float}
\usepackage{multirow}

\usepackage[colorlinks,urlcolor=goodgreen,citecolor=blue,linkcolor=goodred]{hyperref}
\usepackage{bm}
\usepackage{bbm}
\allowdisplaybreaks
\begin{document}
\title{Fabry-P\'erot oscillations in correlated carbon nanotubes}
\author{W. Yang}
\affiliation{ICFO - Institut De Ciencies Fotoniques, The Barcelona Institute of Science and Technology, 08860 Castelldefels (Barcelona), Spain}
\affiliation{Beijing National Laboratory for Condensed Matter Physics and Institute of Physics, Chinese Academy of Sciences, Beijing 100190, PR China}
\author{C. Urgell}
\affiliation{ICFO - Institut De Ciencies Fotoniques, The Barcelona Institute of Science and Technology, 08860 Castelldefels (Barcelona), Spain}
\author{S. L. De Bonis}
\affiliation{ICFO - Institut De Ciencies Fotoniques, The Barcelona Institute of Science and Technology, 08860 Castelldefels (Barcelona), Spain}
\author{M. Marga\'nska}
\affiliation{Institut f\"ur Theoretische Physik, Universit\"at Regensburg, D-93040 Regensburg, Germany}
\author{M. Grifoni}
\affiliation{Institut f\"ur Theoretische Physik, Universit\"at Regensburg, D-93040 Regensburg, Germany}
	\author{A. Bachtold}
	\affiliation{ICFO - Institut De Ciencies Fotoniques, The Barcelona Institute of Science and Technology, 08860 Castelldefels (Barcelona), Spain}
\date{\today}

\begin{abstract}
We report the observation of an intriguing behaviour in the transport properties of nanodevices operating in a regime between the  Fabry-P\'erot and the Kondo limits. Using ultra-high quality nanotube devices, we study how the conductance oscillates when sweeping the gate voltage. Surprisingly, we observe a four-fold enhancement of the oscillation period upon decreasing temperature, signaling a crossover from single-electron tunneling to Fabry-P\'erot interference. These results suggest that the Fabry-P\'erot interference occurs in a regime where electrons are correlated. The link between the measured correlated Fabry-P\'erot oscillations and the SU(4) Kondo effect is discussed.
\end{abstract}

\maketitle

%\begin{multicols}{2}
Electron interactions and quantum interference are central in mesoscopic devices.  The former are due to the electronic charge and give rise to many-body effects; the latter emerges due to the wave-like properties of an electron.  Resonant ballistic devices with a few conduction modes and moderate coupling to electrodes are sensitive to both of these electronic properties.
On the one hand, quantum interference between electron waves backscattered at the boundaries between the mesoscopic system and the metallic electrodes  gives rise to resonant features in the transmission, analogous to the light transmission in an optical Fabry-P\'erot cavity \cite{Datta95}. On the other hand, if the electron spends enough time in the mesoscopic device before being transmitted, Coulomb repulsion can also become important giving rise to Coulomb blockade and single-charge tunneling effects \cite{Kouwenhoven97}. Despite considerable efforts, the interplay between electron interactions and quantum interference remains poorly understood from both an experimental and a theoretical point of view, due to the many-body character of the problem. This is the topic of the present Letter.

Carbon nanotubes (CNTs), semiconducting nanowires, and edge channels of the quantum Hall effect are ideal quasi one-dimensional (1d) systems to study both electron correlations and quantum interference. In fact, various many-body effects including Coulomb blockade~\cite{Tans97,Bockrath97,DeFranceschi2003}, Wigner phases \cite{Deshpande2008,Pecker2013,Shapir2019,Lotfizadeh2019}, and  Kondo physics~\cite{Nygard2000,Jarillo-Herrero2005,Paaske2006,Makarovski2007PRL,Fang2008,Cleuziou2013,Schmid2015,Niklas2016,Ferrier2016,Ferrier2017,Desjardins2017,Jespersen2006} as well as Fabry-P\'erot and Mach-Zehnder oscillations  resulting from electron interference
 \cite{Liang2001,Litvin2007,Kim2007,Dirnaichner2016,Lotfizadeh2018,Kretinin2010,Neder2006} have been observed in these multi-mode 1d systems.
It is possible to switch from interaction- to interference-governed transport regimes by tuning the tunnel couplings at the interface between the wire and the electrodes, $\Gamma_{S}$ and $\Gamma_{D}$ for the source $(S)$ and drain $(D)$ electrodes. Which transport regime dominates crucially depends on how large the tunneling broadening $\hbar \Gamma=\hbar(\Gamma_{S}+\Gamma_{D})$ is compared to other energy scales, in particular to the charging energy $E_C$, being the electrostatic cost to add another (charged) electron to the wire \cite{Laird2015}. In the so-called quantum dot limit, characterized by  $\hbar \Gamma \ll E_C$, tunneling events in and out of the wire are rare and Coulomb charging effects are dominant. They give rise to Coulomb blockade phenomena and incoherent single-electron tunneling in the regime $\hbar \Gamma <k_BT \ll E_C $. By decreasing temperature, one expects coherent single electron tunneling for $k_BT \simeq\hbar \Gamma \ll E_C$, where the width of the Coulomb peaks is determined by $\Gamma$;  at even lower temperatures, when spin-fluctuations become relevant, the Kondo effect emerges as dominant transport mechanism. In the opposite limit of large transmission, $\hbar \Gamma \gg E_C$,
interference effects give rise to the characteristic
Fabry-P\'erot
 patterns, which can be easily calculated from a non-interacting single-particle scattering approach \cite{Liang2001}.
    In the focus of this Letter is the intermediate transmission regime $\hbar\Gamma\sim E_C\gg k_BT$ when no clear hierarchy of energy scales exists.

\begin{figure}[tb!]
	\begin{center}
		\includegraphics[width=\linewidth]{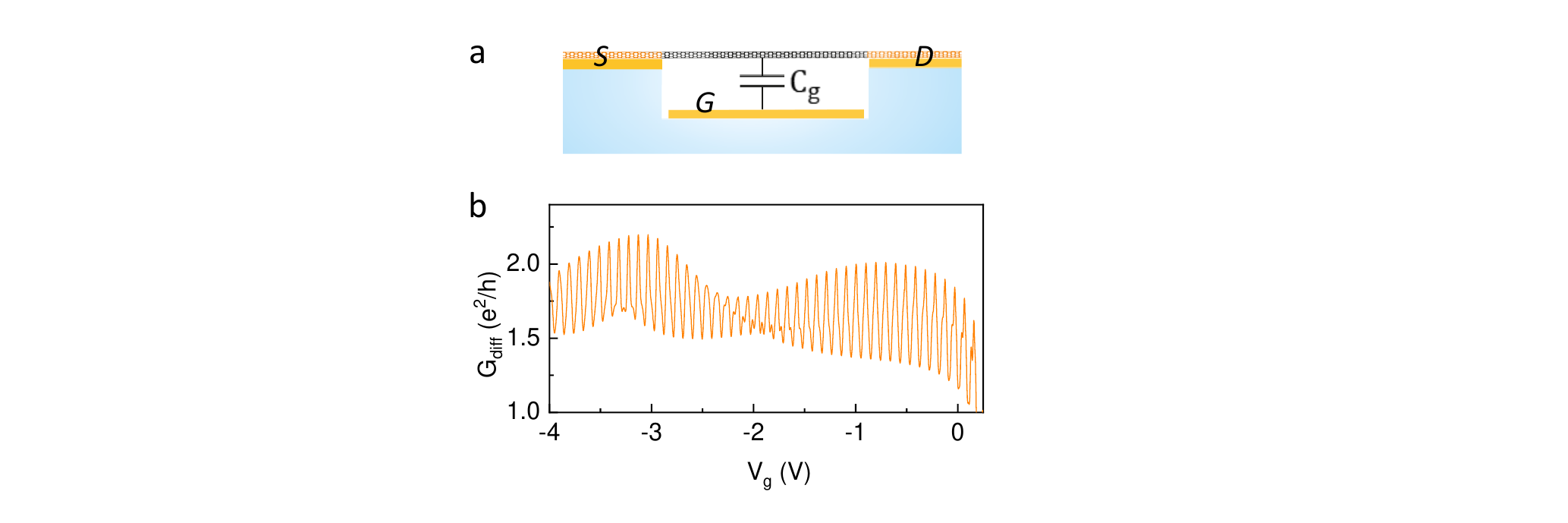}
		\caption{ Schematics of the device and low-temperature  transport characteristics. (a) Three-terminal device with a suspended CNT contacted to  source ($S$), drain ($D$), and gate ($G$) electrodes.  (b) Gate voltage dependence of the conductance at zero source-drain voltage of device I at $T$=15 mK. An oscillating voltage with amplitude smaller than $k_BT/e$ is applied to measure the differential conductance.}
		\label{fig1}
	\end{center}
\end{figure}

An experimental hallmark of both interaction- and interference-dominated transport is the modulation of the conductance when sweeping the electrochemical potential, that is, by varying the gate voltage $V_{g}$.  In the incoherent tunneling regime, the alternance of single-electron tunneling and  Coulomb blockade physics results in finite conductance peaks with a period in $V_{g}$ of the order of $e/C_{g}$ \cite{Kouwenhoven97}, where $-e$ is the (negative) electron charge and $C_{g}$ is the capacitance between the nanotube and the gate electrode, see Fig.~\ref{fig1}(a).
 In contrast, in the interference-dominated regime  the conductance modulation of the Fabry-P\'erot oscillations arises from the electron wave phase accumulated during a round trip along the wire.
 The presence of valley and spin degrees of freedom in CNTs gives rise to interferometers with oscillation period $\Delta^{} V_{g} =4e/C_g$ \cite{Liang2001}.

In this work, we improve the quality {of nanotube devices} to an unprecedented level. We discover a crossover of the conductance oscillation period between $e/C_{g}$ and $4e/C_g$ upon sweeping temperature. Above liquid helium temperature, the period is $e/C_{g}$ with oscillations amplitudes pointing to coherent single-electron tunneling in an open quantum dot configuration. At low temperature, the period becomes $4e/C_g$ and the oscillations feature typical characteristics of Fabry-P\'erot interference. These unexpected data are a clear signature of the interplay between interaction and quantum interference.

\textit{Experimental results.-} We grow nanotubes by chemical vapor deposition on prepatterned electrodes \cite{Cao2005}. The nanotube is suspended between two metal electrodes, see Fig.~\ref{fig1}(a). We clean the nanotube in the dilution fridge at base temperature by applying a high constant source-drain voltage $V_{\rm sd}$ for a few minutes (see Sec. I of the Supplemental Material). This current-annealing step cleans the nanotube surface from contamination molecules adsorbed when the device is in contact with air.
%The value of $V_{\rm sd}$ is chosen by ramping up the bias until the point when the current starts to decrease, see Fig.~\ref{fig1}(b). This current-annealing step cleans the nanotube surface from contaminations. This procedure allows us to adsorb helium monolayers uniformly along nanotubes, indicating that the nanotubes are essentially free of adsorbate contamination \cite{Noury2019}.
The energy gap of the two nanotubes discussed in this work is on the order of 10 meV (for details
see the Supplemental Material). The length of the two suspended nanotubes inferred by scanning electron microscopy (SEM) is about 1.5 $\mu$m.

Figure~\ref{fig1}(b) shows the modulation of the differential conductance $G_{\rm diff}$ of device I as a function of $V_{g}$ in the hole-side regime at 15 mK. Rapid conductance oscillations are superimposed on slow modulations. Since the conductance remains always large, that is above $e^2/h$, we attribute the rapid oscillation to the Fabry-P\'erot interference with period in gate voltage being  $\Delta V_g= 4e/C_{g}$.
The slow modulation may be caused by the Sagnac interference \cite{Dirnaichner2016,Lotfizadeh2018}, the additional backscattering due to a few residual adatoms on the CNT, the symmetry breaking of the electronic wave function by the planar contacts of the device, or any combination of these (for further discussion see Sec. I and IIA  of the Supplemental Material).
%The pattern of the secondary interference is completely changed each time that we do a current-annealing of the device (section II of Supplemental Material). We attribute this modification to the atomic rearrangement of the platinum electrodes in the region near the nanotube, so that the intervalley backscattering rate at the contacts changes \cite{Dirnaichner2016}. Before any current annealing, the conductance modulation appears much less regular, see Fig. \ref{fig1}(c).
%
\begin{figure*} %[htb!]
	\begin{center}
	\includegraphics[width=1.0\textwidth]{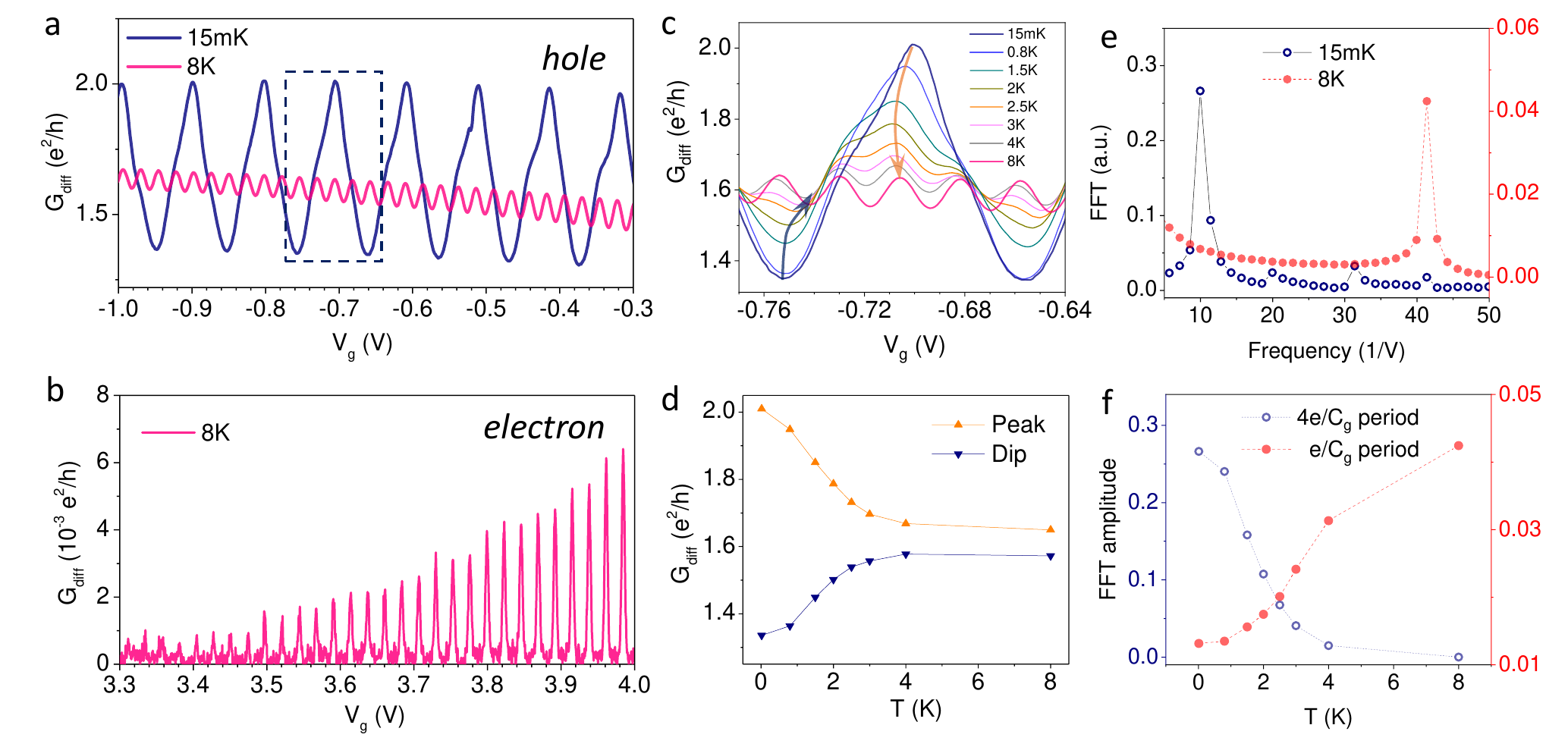}
		\caption{ Temperature-induced crossover from an interference-dominated to a charging-controlled regime in device I. (a,b) Oscillations of the conductance $G_{\rm diff}(V_{g})$ versus gate voltage $V_{g}$ in the hole- and electron-doped regimes. (c) Evolution of the oscillation period for a series of different temperatures. The range of $V_{g}$ shown in this figure is highlighted in panel (a) by a dashed rectangle. (d) Temperature dependence of the conductance associated with a peak and a dip, as indicated by arrows in  (c). (e) Fast Fourier transform (FFT) of the $G_{\rm diff}(V_{g})$ traces at 15 mK and 8 K measured for $V_{g}$ between -1.0 V and -0.3 V. (f) Temperature dependence of the FFT amplitude associated with the $4e/C_{g}$  period oscillations and the $e/C_{g}$  period oscillations. }
		\label{fig2}
	\end{center}
\end{figure*}

\begin{figure}[htb]
	\begin{center}
	\includegraphics[width=\linewidth]{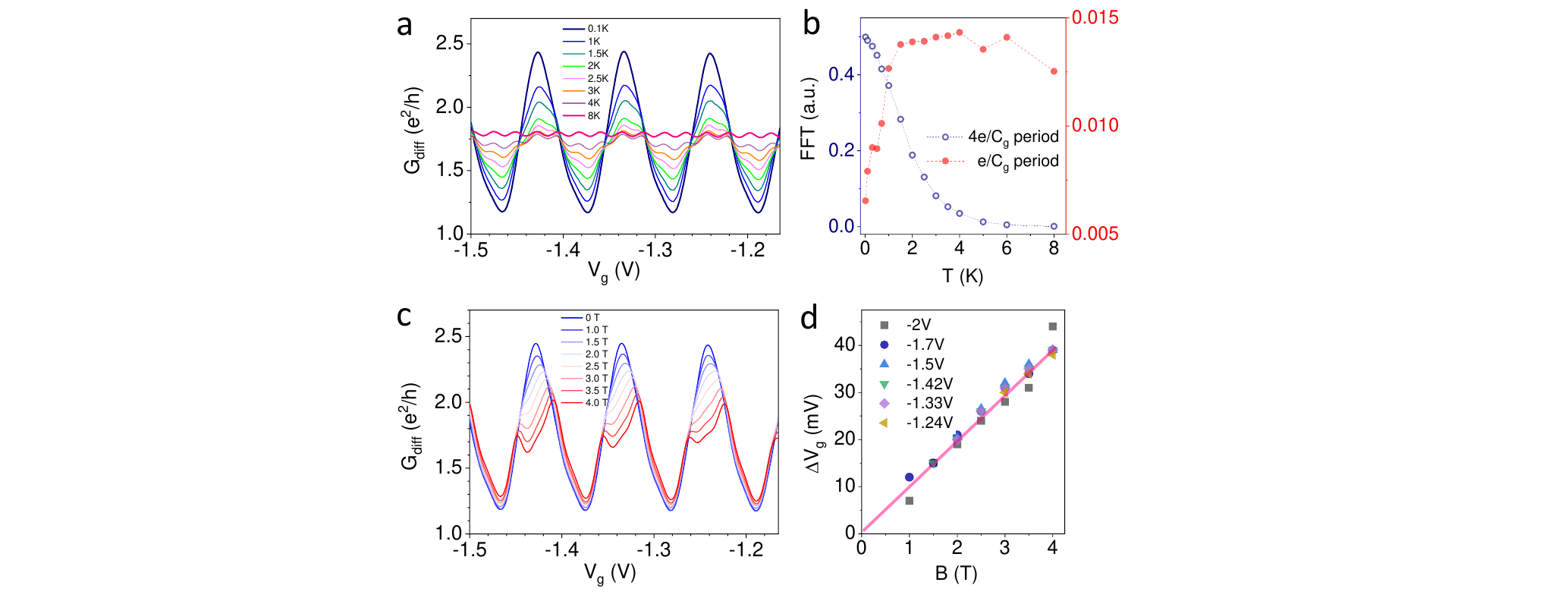}
		\caption{Measurements on device II. (a) Conductance traces for a series of different temperatures. (b) Temperature dependence of the FFT amplitude associated with the $4e/C_{g}$  and the $e/C_{g}$  period oscillations. (c) Conductance traces for different perpendicular magnetic fields at 15 mK. (d) Peak splitting as a function of magnetic field for the conductance peaks at different gate voltages.}
		\label{fig3}
	\end{center}
	\end{figure}

A crossover to a regime dominated by the charging effect in an open interacting quantum dot is observed upon increasing temperature.
Specifically, by sweeping the temperature from 15 mK to 8 K the amplitude of the oscillations gets smaller. Further,
the oscillation period gets four times lower, changing from $4e/C_{g}$ at 15 mK to $e/C_{g}$ at 8~K, see Figs.~\ref{fig2}(a) and (c-e).
The period in $V_{g}$ is calibrated in units of $e/C_{g}$ using the measurements in the electron-side regime, where regular Coulomb oscillations are observed at 8~K, as shown in Fig.~\ref{fig2}(b).
The same behavior is observed in device II, Figs.~\ref{fig3}(a) and (b). The $4e/C_{g}$ oscillations vanish at $\sim 3$~K in both devices, whereas the $e/C_{g}$ oscillation amplitude is suppressed to almost zero below $\sim 1$ K in device I and below $\sim 0.1$ K in device II, see Figs.~\ref{fig2}(f) and \ref{fig3}(b).

\begin{figure}[htb]
	\begin{center}
		\includegraphics[width=\linewidth]{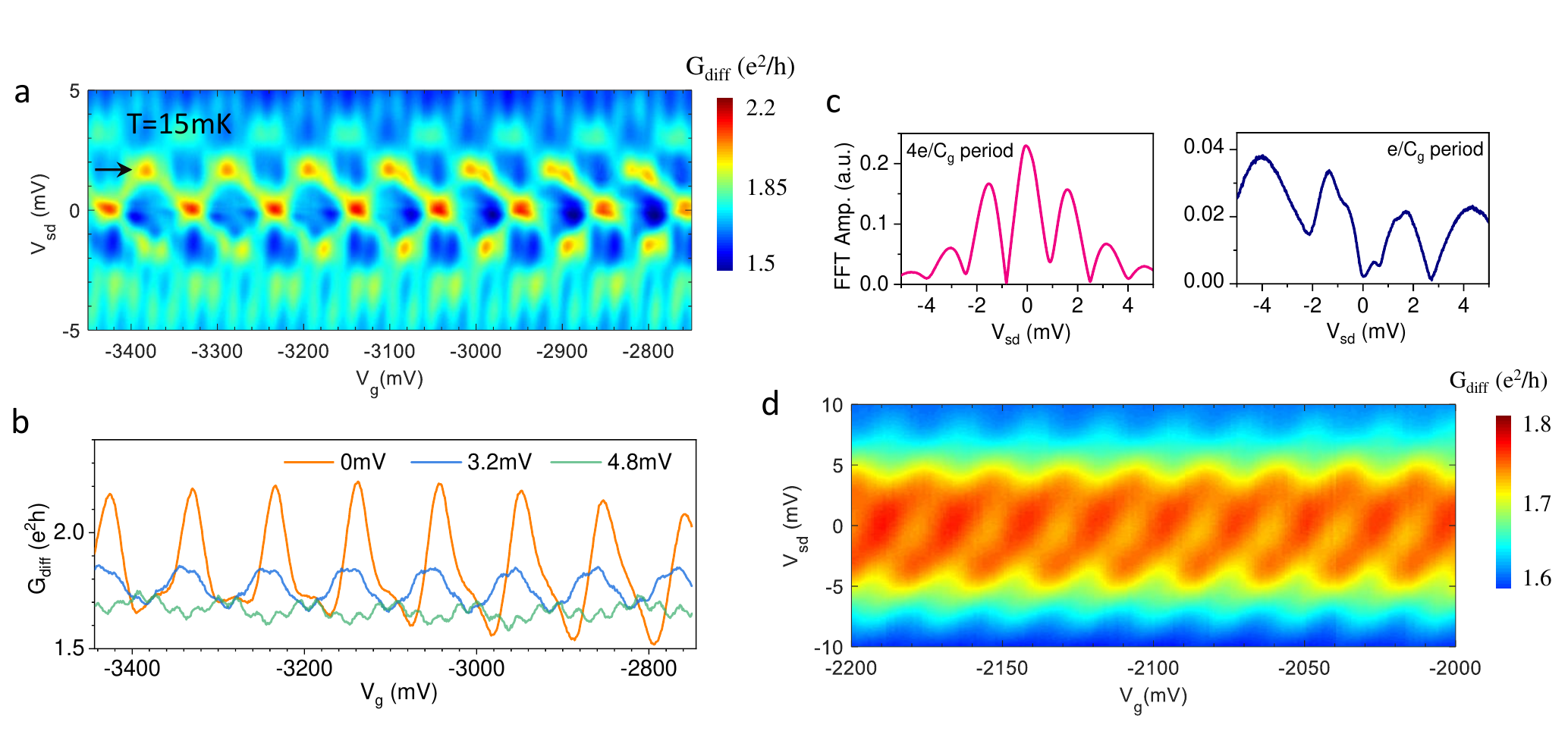}
		\caption{From Fabry-P\'erot patterns to blurred Coulomb diamonds in device I. (a) Map of the differential conductance as a function of $V_{\rm sd}$ and $V_{g}$ at 15 mK. From the position of the arrow the single-particle excitation energy is extracted. (b) Differential conductance traces for a series of different source-drain voltages at 15 mK. (c) Source-drain voltage dependence of the FFT amplitude associated with the $4e/C_{g}$ and the $e/C_{g}$  period oscillations at 15~mK. The curves are obtained by doing a FFT of the $G_{\rm diff}(V_{g})$ trace for each $V_{\rm sd}$ value. (d) Map of the differential conductance as a function of $V_{\rm sd}$ and $V_{g}$ at 8 K. The dashed lines highlight the contours of the Coulomb diamonds.}
		\label{fig4}
	\end{center}
\end{figure}

Our interpretation of a temperature-induced crossover between two seemingly distinct transport regimes is confirmed by measured maps of the differential conductance as a function of source-drain and gate voltages at $T$=15~mK and $T$=8~K, as shown in Fig.~\ref{fig4}(a) and (d), respectively. The low-temperature data feature the regular chess-board-like Fabry-P\'erot interference pattern  \cite{Liang2001}, while the high-temperature data show smeared Coulomb diamonds. Such measurements further allow us to extract important energy scales for our device. The characteristic bias $V_{\rm sd}^*$ indicated by the arrow in Fig.~\ref{fig4}(a) yields a single-particle excitation energy $\Delta E=eV_{\rm sd}^{*}\simeq 1.7$ meV. This value is consistent with what is expected from a nanotube with length $L\simeq$ 1.5 $\mu$m. Assuming the linear dispersion $\varepsilon(k)=\hbar v_F k$ with longitudinal quantization $k_n=n\pi/L$ and the Fermi velocity $v_F=10^6$~m/s, it yields $\Delta E=\varepsilon (k_{n+1})-\varepsilon(k_n)=\hbar v_F \pi/L \simeq 1.4$ meV.
The charging energy is estimated from the charge stability diagram measurements at 8 K, Fig.~\ref{fig4}(d); from the Coulomb diamond, indicated by the dashed lines, a charging energy  $E_C\simeq 3.6$ meV is extracted. Further, we estimate $\hbar \Gamma \sim E_C$ because of the strong smearing of the diamonds in Fig.~\ref{fig4}(d) and the weak conductance modulation at 8~K in Fig.~\ref{fig2}(a).
%\cite{Joyez1997}.
The energy hierarchy in our experiment is thus $E_C \simeq \hbar \Gamma  \simeq \Delta E\gg k_B T$.

%\begin{figure*}[htb!]
%\end{figure*}

The evolution of the 15~mK conductance oscillations as a function of the source-drain bias shows that both oscillations {\em{coexist}} over a large bias range, albeit with modulated strengths, see Figs.~\ref{fig4}(a-c). The main trend is that the oscillation period changes from $4e/C_{g}$ at zero bias to $e/C_{g}$ at high bias. By contrast, the evolution in perpendicular magnetic field shows that the conductance peaks are split in two, with the splitting in gate voltage being linear in magnetic field, see Figs.~\ref{fig3}(c,d). This is attributed to the Zeeman spitting, since the associated $g$-factor is $2.4\pm0.4$. The error in the estimation arises from the uncertainty in the lever arm. These data indicate degeneracy of the four electron levels associated to the spin and valley degrees of freedom.

\textit{Discussion.-}
We examine possible origins of the temperature-induced period change. Let us first assume that interactions are not important. Then, upon lowering temperature, noninteracting Fabry-P\'erot oscillations are expected to emerge when the thermal smearing becomes smaller than the single-particle excitation energy. However,  thermal smearing is associated to a characteristic temperature $T_{th} \sim \Delta E/k_B \approx 20$~K, which is rather different from the measured crossover temperature $T_C\sim 3$~K in Fig~\ref{fig2}(f) and \ref{fig3}(c). In addition, thermal smearing cannot explain the emergence at temperatures above $T_C$ of the $e/C_{g}$ oscillations due to coherent single-electron tunneling. Therefore, thermal decoherence is not at the origin of the measured period change. This is further supported by single-particle Fabry-P\'erot interference calculations, based on an accurate tight-binding modeling of CNTs,  that we carried out. We also considered the complementary regime, and investigated whether charge fluctuations could be the cause of our finding. However, when using an interacting multilevel quantum dot with four-fold degenerate energy levels in the regime $E_C \simeq\hbar \Gamma$, we could not reproduce the measured fourfold variation of the period. Both the single-particle and the interacting calculations are described in the Sec. II of the Supplemental Material.

The high-temperature measurement of the charging effect in an open quantum dot indicates electron correlation. When reducing temperature, the associated $e/C_{g}$ conductance oscillations disappear smoothly to give rise to the $4e/C_{g}$ oscillations. The smoothness of the crossover suggests that the Fabry-P\'erot-like oscillations also occur in a regime where electrons are correlated.
This smooth change of periodicity bears similarities but also differences compared to the SU(4) Kondo effect in carbon nanotubes, occurring in the weak tunneling regime $E_C \gg\hbar \Gamma >k_BT $ \cite{Ferrier2016,Anders2008}. In the Kondo effect, the tunneling coupling is low enough compared to the charging energy to allow full localization of the charge within the dot, but it is large enough compared to the Kondo energy to enable both spin and valley fluctuations~\cite{Jarillo-Herrero2005}. This results in a crossover from charging effects at high temperature to the increased conductance of Kondo resonances at zero temperature, with a fourfold enhancement of the oscillation period~\cite{Laird2015,Makarovski2007PRL,Anders2008}. In contrast to our observations though, in the SU(4) Kondo effect  the conductance alternates between large values close to $4e^2/h$ at oscillation maxima and almost zero at minima \cite{Anders2008,Ferrier2016}; see also Sec. Ib of the Supplemental Material.  In our annealed devices, the tunneling coupling is large, $\hbar \Gamma \simeq E_C$. The charge is no longer strongly localized within the dot. As a result, our devices are in a regime where there are also charge fluctuations in the nanotube, in addition to spin and valley fluctuations. This might be at the origin of the crossover of the conductance oscillation period observed in this work, similar to what happens in
the SU(4) Kondo regime~\cite{Laird2015,Makarovski2007PRL,Anders2008}, but with conductance minima clearly distinct from zero.
We emphasize that the zero-source-drain bias, low-temperature $G_{\rm diff}(V_{g})$ data alone do {\em{not}} allow one to distinguish between non-interacting and correlated Fabry-P\'erot oscillations.  However, the smooth modulation between $e/C_g$ and $4e/C_g$ oscillations upon increasing the bias (see Fig.~\ref{fig4}(c)) further supports our hypothesis of correlated Fabry-P\'erot regime.

\textit{Conclusion.-} Our work provides a comprehensible phenomenology of transport in nanotubes when both interference and interaction are involved. The findings presented in this work have been possible thanks to the high quality of the devices, since otherwise disorder leads to irregular $G_{\rm diff}(V_{g})$ modulations that are difficult to interpret. The main results are summarized as follows: (a) We measure a fourfold enhancement of the oscillation period of $G_{\rm diff}(V_{g})$ upon decreasing temperature, signaling a crossover from coherent single-electron tunneling to Fabry-P\'erot interference; both oscillations coexist at the crossover temperature. (b) Upon increasing the source-drain bias at low temperature, both oscillations coexist over a large bias range. (c) The Sagnac-like modulation pinpoints the quantum interference nature of the Fabry-P\'erot oscillations at zero bias. (d) The magnetic field data suggest a four-fold spin and orbital degeneracy at zero-magnetic field.

The unexpected temperature-induced crossover, possibly related to charge, spin, and valley fluctuations, raises an important question: How does the strength of charge fluctuations compare to that of spin and valley fluctuations in our experiment? Indeed,  when the electron transmission approaches one in open fermion channels, the electron shot noise is suppressed to zero~\cite{Blanter2000}, indicating that there are no longer any charge, spin, and valley fluctuations in nanotubes; by contrast, in the lower $\Gamma$ limit of SU(4) Kondo, spin and valley fluctuate, but not the charge. It is then natural to ask how the crossover temperature in our devices compares with the well-known Kondo temperature of closed quantum dots. However, a quantitative description of our experiment constitutes a theoretical challenge. It will be interesting to measure shot noise \cite{Onac2006,Recher2006,Wu2007,Herrmann2007} and the backaction of the electro-mechanical coupling \cite{Urgell2020,Wen2020} to further characterize these correlated Fabry-P\'erot oscillations.

We thank B. Thibeault at UCSB for fabrication help, W.J. Liang, P. Recher and F. Dolcini for discussions. This work is supported by ERC advanced grant number 692876, the Cellex Foundation, the CERCA Programme, AGAUR (grant number 2017SGR1664), Severo Ochoa (grant number SEV-2015-0522), MICINN grant number RTI2018-097953-B-I00 and the Fondo Europeo de Desarrollo Regional. We acknowledge support by the Deutsche Forschungsgemeinschaft within SFB 1277 B04.
%\bibliographystyle{apsrev}
%\bibliography{biblPRLmain}

%\end{multicols}

\newpage

\begin{center}
	\bf{Fabry-P\'erot oscillations in correlated carbon nanotubes
		\\Supplemental Material}
\end{center}

\section{Experimental section}
\subsection{High-quality nanotubes obtained by current annealing}
\begin{figure*}[tb!]
	\begin{center}
		\includegraphics[width=17cm]{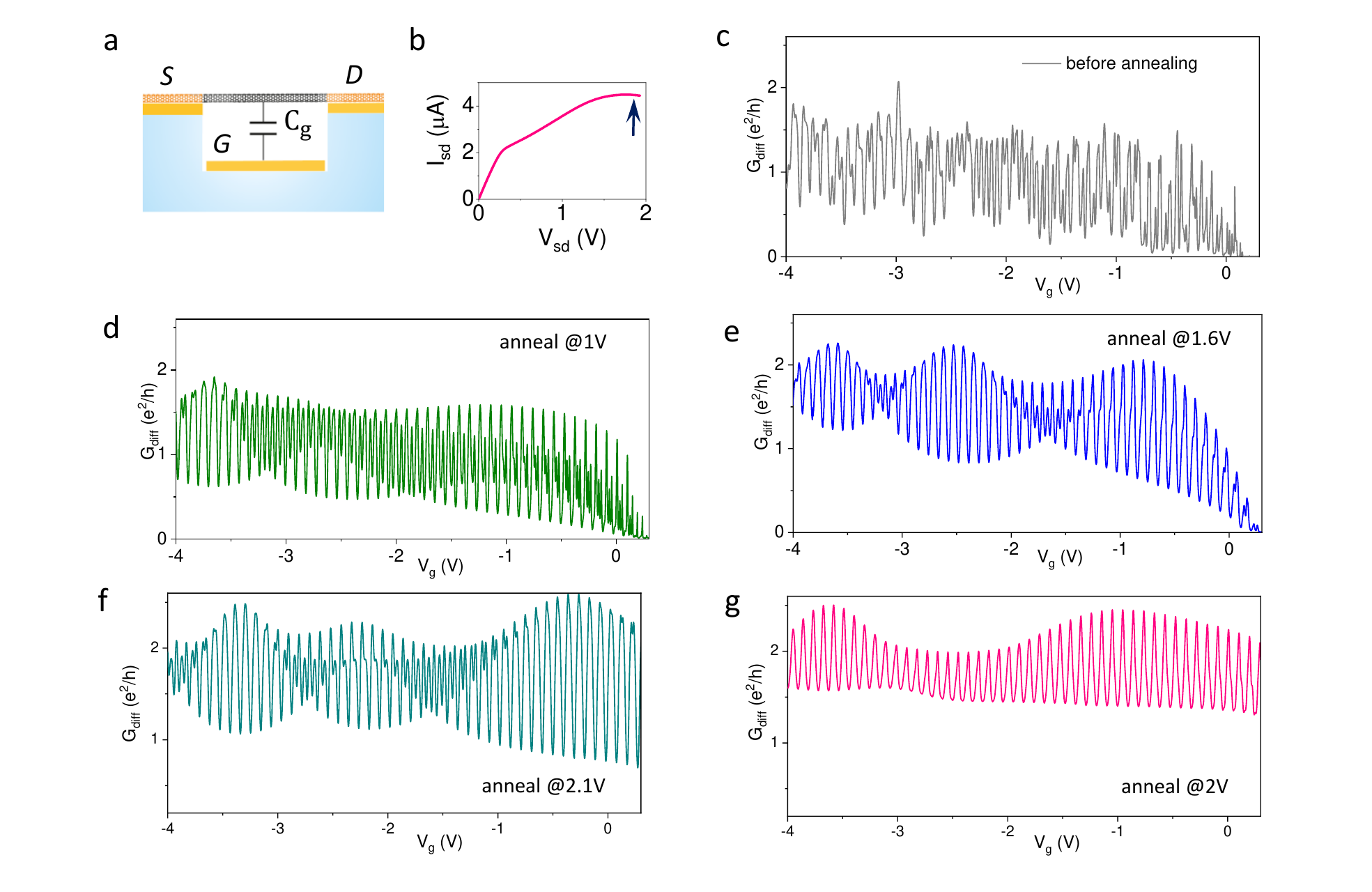}
		\caption{Current annealing and low-temperature  transport characteristics. (a) Three-terminal device with a suspended CNT contacted to  source (S), drain (D), and gate (G) electrodes. (b) Current-voltage characteristic of device I at $T$=15 mK. The arrow indicates when the current starts to decrease while increasing  $V_{\rm g}$. The highest voltage used for current annealing is usually around this value. (c-g) Gate voltage dependence of the conductance $G_{\rm diff}(V_{\rm g})$  of device I at $T$=15 mK measured before current annealing and after different current annealing steps. The measurements in d-g have been carried out in a second cool-down, while all the other presented data of device I have been recorded in the first cool-down. An oscillating voltage with amplitude smaller than $k_BT/e$ is applied to measure the differential conductance.}
		\label{figS1}
	\end{center}
\end{figure*}

We grow nanotubes by chemical vapor deposition on prepatterned electrodes using the technique described in Ref.~[\onlinecite{Cao2005}]. The nanotube is suspended between two metal electrodes Fig.~\ref{figS1}(a).  We clean the nanotube in the dilution fridge at base temperature by applying a high constant source-drain voltage $V_{\rm sd}$ for a few minutes. The highest applied value of $V_{\rm sd}$ is usually chosen by ramping up the bias until the point when the current starts to decrease, see Fig.~\ref{figS1}(b). This current-annealing step cleans the nanotube surface from contaminations. This procedure allows us to adsorb helium monolayers uniformly along nanotubes, indicating that the nanotube is essentially free of adsorbate contamination \cite{Noury2019}. Figures~\ref{figS1}(c,g) show the modulation of the differential conductance $G_{\rm diff}$ of device I as a function of $V_{\rm g}$ in the hole-side regime at 15 mK before and after annealing, respectively. The current annealing results in regular conductance modulation.

\begin{figure}[tb!]
	\begin{center}
		\includegraphics[width=9cm]{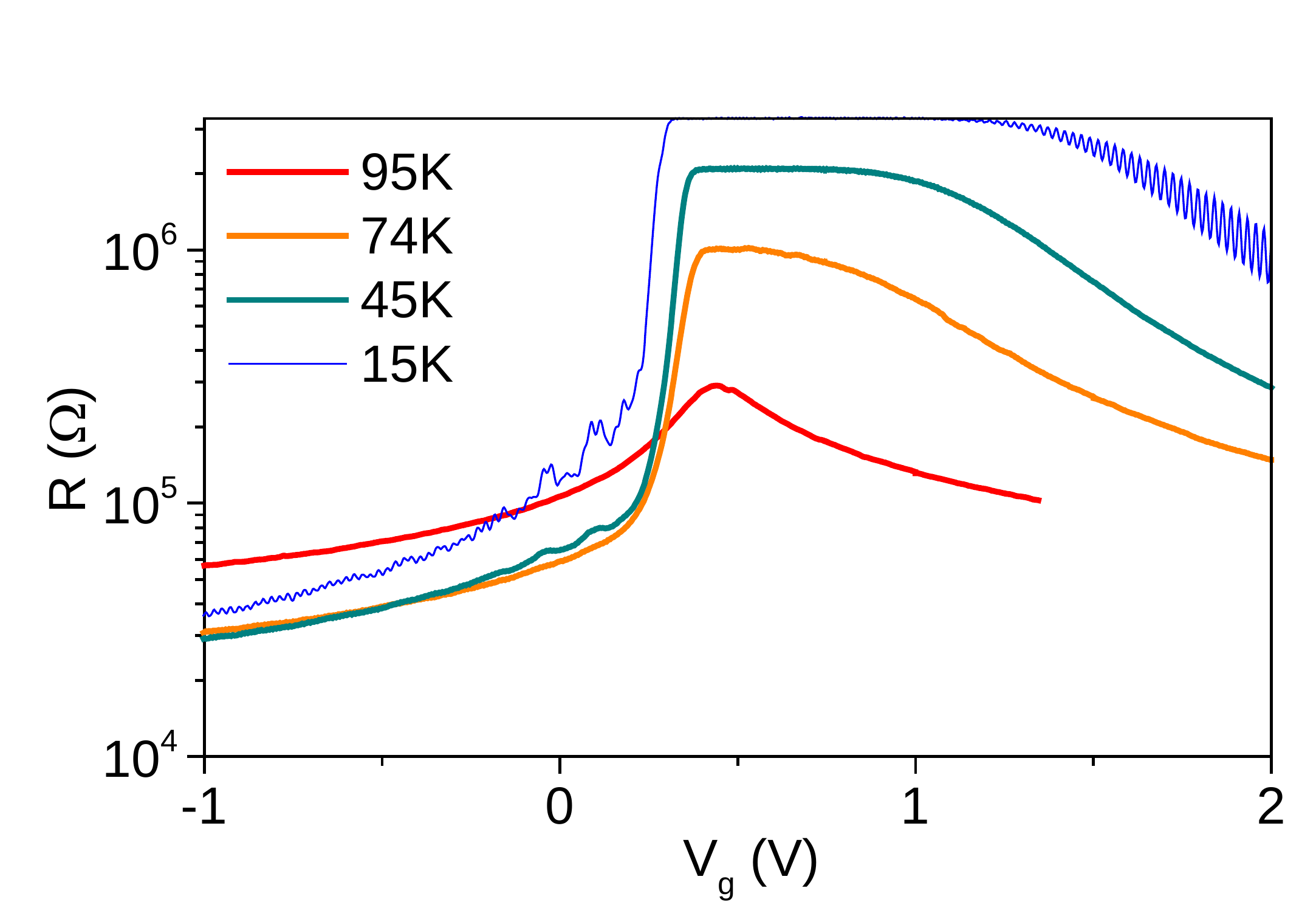}
		\caption{Resistance of device I as a function of gate voltage for different temperatures. }
		\label{figS2}
	\end{center}
\end{figure}

In the annealed sample rapid conductance oscillations are superposed on slow modulations, see Fig.~\ref{figS1}(d). Since the conductance remains always large, we attribute the rapid oscillation to the Fabry-P\'erot interference with period in gate voltage being  $\Delta V_{\rm g}= 4e/C_{\rm g}$. The first interpretation of slow modulation coming to mind is the so-called Sagnac interference, due to the gradual change of the Fermi velocity when sweeping $V_{\rm g}$,\cite{Dirnaichner2016,Lotfizadeh2018}, caused by the trigonal warping. In the dispersion of non-interacting electrons trigonal warping manifests at energies further than $\sim200$~meV away from the charge neutrality point, while the range of single-particle energies scanned in our experiment is of the order of $\sim56$~meV (estimated from $\sim40$ peaks visible in Fig. 1(b) of the main text, separated by $\Delta E \simeq 1.4$~meV). Unless the interactions bring the trigonal warping effects closer to the charge neutrality point, an alternative explanation of the slow modulation is needed. One possibility is the beating caused by the presence of a symmetry breaking mechanism which introduces additional valley mixing and/or another characteristic length scale into the system (see the discussion of Fig.~\ref{suppfig:FPbroken}). The pattern of the secondary interference is completely changed each time that we do a current-annealing of the device, see Fig.~\ref{figS1}(d,e). We attribute this modification either to the atomic rearrangement of the platinum electrodes in the region near the nanotube, so that the intervalley backscattering rate at the contacts changes \cite{Dirnaichner2016}, or to the changed position of residual adatoms near the contacts.

The effect of the annealing on device II are discussed in the next subsection, see Fig.~\ref{figS4}.
\subsection{Electron transport properties}
In this subsection we provide additional data to further characterize device I and II discussed in the main text.

{\em{Size of the energy gap}}- The energy gap of the two nanotubes discussed in this work is on the order of 10 meV. The size of the energy gap can be obtained by recording the dependence of the resistance on $V_{\rm g}$ at different temperatures \cite{Laird2015}, see Fig.~\ref{figS2}. The order of magnitude of the band gap $E_{\rm G}$ is obtained from the temperature at which the resistance in the gap gets high, $E_{\rm G}\sim k_{\rm B}T$.

\begin{figure*}[tb!]
	\begin{center}
		\includegraphics[width=\textwidth]{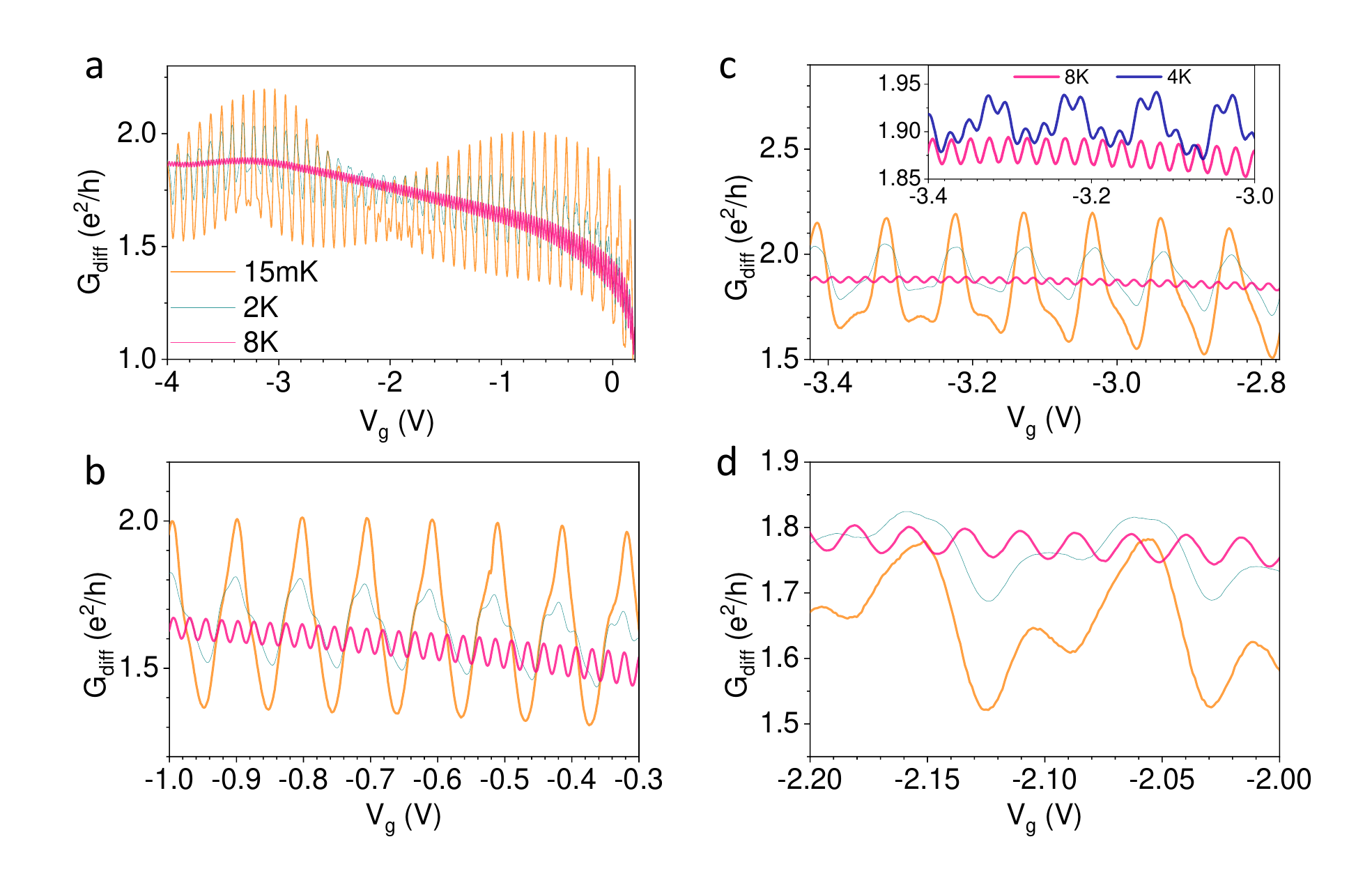}
		\caption{Series of $G_{\rm diff}(V_{\rm g})$ traces at different temperatures of device I. We select the $V_{\rm g}$ ranges for which data are presented in the main text.}
		\label{figS3}
	\end{center}
\end{figure*}

{\em{Temperature dependence of device I}}- In Fig.~\ref{figS3} is shown a selection of $G_{\rm diff}(V_{\rm g})$ traces of device I at different temperatures. We select the $V_{\rm g}$ ranges for which data are presented in the main text. The change in period of the oscillations with temperature is observed for all the gate voltage ranges.

\begin{figure*}[tb!]
	\begin{center}
		\includegraphics[width=16cm]{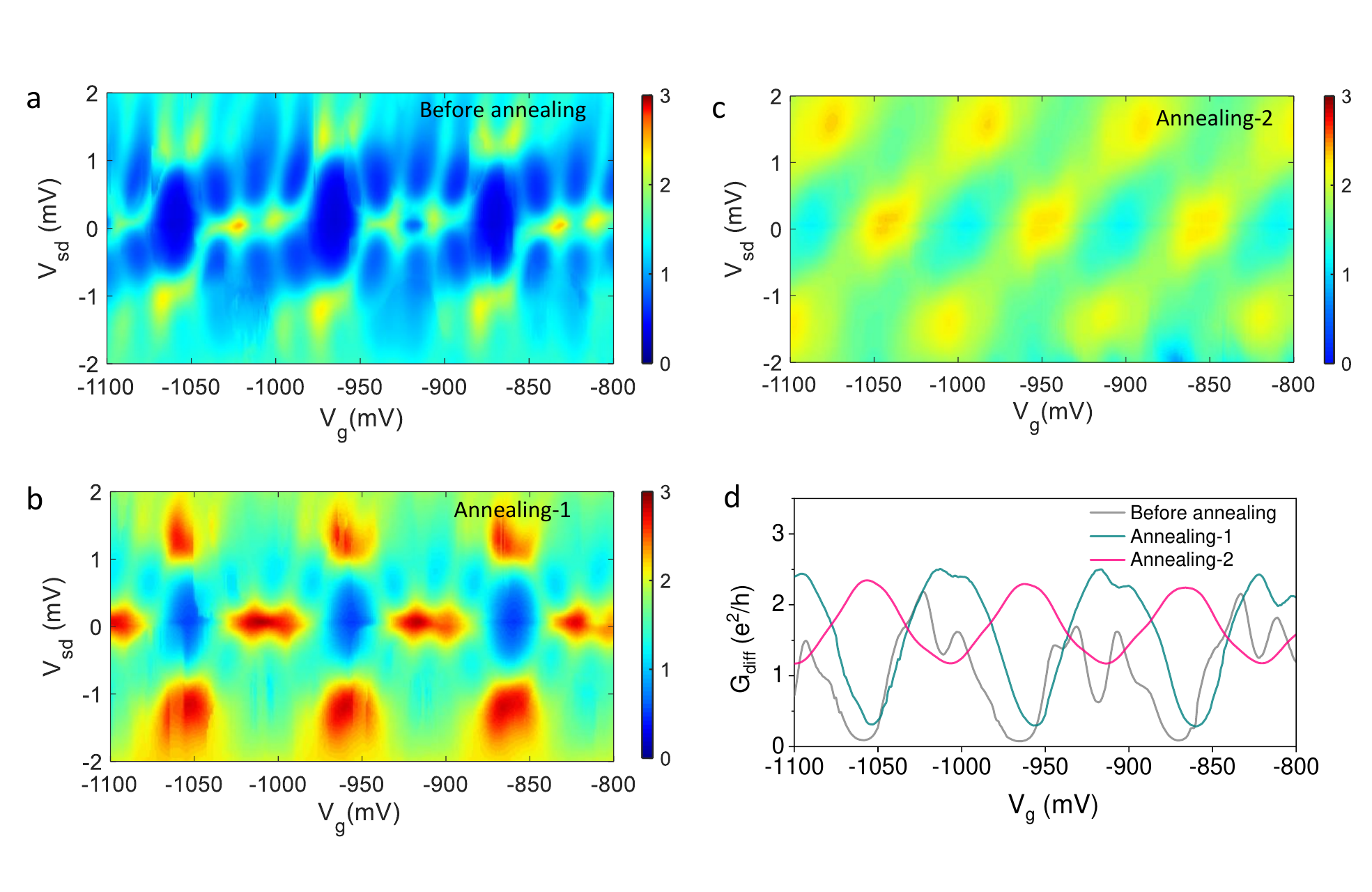}
		\caption{Differential conductance as a function of $V_{\rm sd}$ and $V_{\rm g}$ of device II after various annealing steps at $T$=15mK. Before annealing, panel (a), the coupling to leads is such that clear Kondo ridges are observed at low bias. Such features survive  after the first annealing step, as seen in panel (b). After the second annealing step, panel (c), the differential conductance is overall larger and Fabry-P\'erot features are seen. Conductance traces are compared in panel (d). }
		\label{figS4}
	\end{center}
\end{figure*}

{\em{Change from intermediate to strong coupling upon annealing }}- Finally, we show in Fig.~\ref{figS4} the effect of successive annealing steps on the map of the differential conductance as a function of $V_{\rm sd}$ and $V_{\rm g}$ of device II. Remarkably, before annealing regions of very low differential conductance alternate with regions of high conductance in a way which is reminiscent of the SU(2)xSU(2) Kondo effect seen in other CNT-based quantum dots \cite{Laird2015}. Here, as seen from the conductance trace in Fig.~\ref{figS4}(d), within a periodicity of four electrons, an enhancement of the conductance is seen in the odd valleys. After the first annealing, a stronger coupling to the leads favours a conductance enhancement also in the intermediate valley, a signature of the formation of an SU(4) Kondo state.
The second annealing leads to an even larger coupling to the leads, and the Kondo features are no longer seen at low bias. Rather, a checkerboard pattern typical of Fabry-P\'erot interference is the dominant feature.

\section{Theoretical calculation of transport}

Because of the lack of clear energy scales separation, i.e. $U \simeq \Gamma \gg k_BT$, the theoretical description reproducing the results of the experiments is very challenging; $U=E_C$ stands for the characteristic strength of the Coulomb interaction between the electrons in the system. We can however provide theoretical support for our interpretation of the data as the interplay of correlations and interference effects by showing that neither of these mechanisms alone can explain the observed evolution of conductance with temperature. On one hand, we show in Sec. II A results for the Fabry-P\'{e}rot interference with $\Gamma\gg k_BT$ and  $U=0$. While such single-particle interference can explain the experimental results at 15 mK, it cannot reproduce the fourfold decrease in the oscillation period with increasing temperature. On the other hand, we analyze in Sec. II B the electronic transport across an interacting multilevel quantum dot with  four-fold degenerate energy levels and level spacing $\Delta E$. We use a so-called coherent sequential tunneling approximation, which yields correct results for non-interacting ($U=0$) and Coulomb blocked ($U\gg k_BT>\Gamma$) systems, but also in the regime $U > \Gamma \gtrsim k_BT$. Lowering the temperature again does not introduce any change in periodicity. An essential ingredient, the Kondo-like correlation, is missing from the theory.

\subsection{Single particle Fabry-P\'{e}rot interference}
\begin{figure*}
 \includegraphics[width=\textwidth]{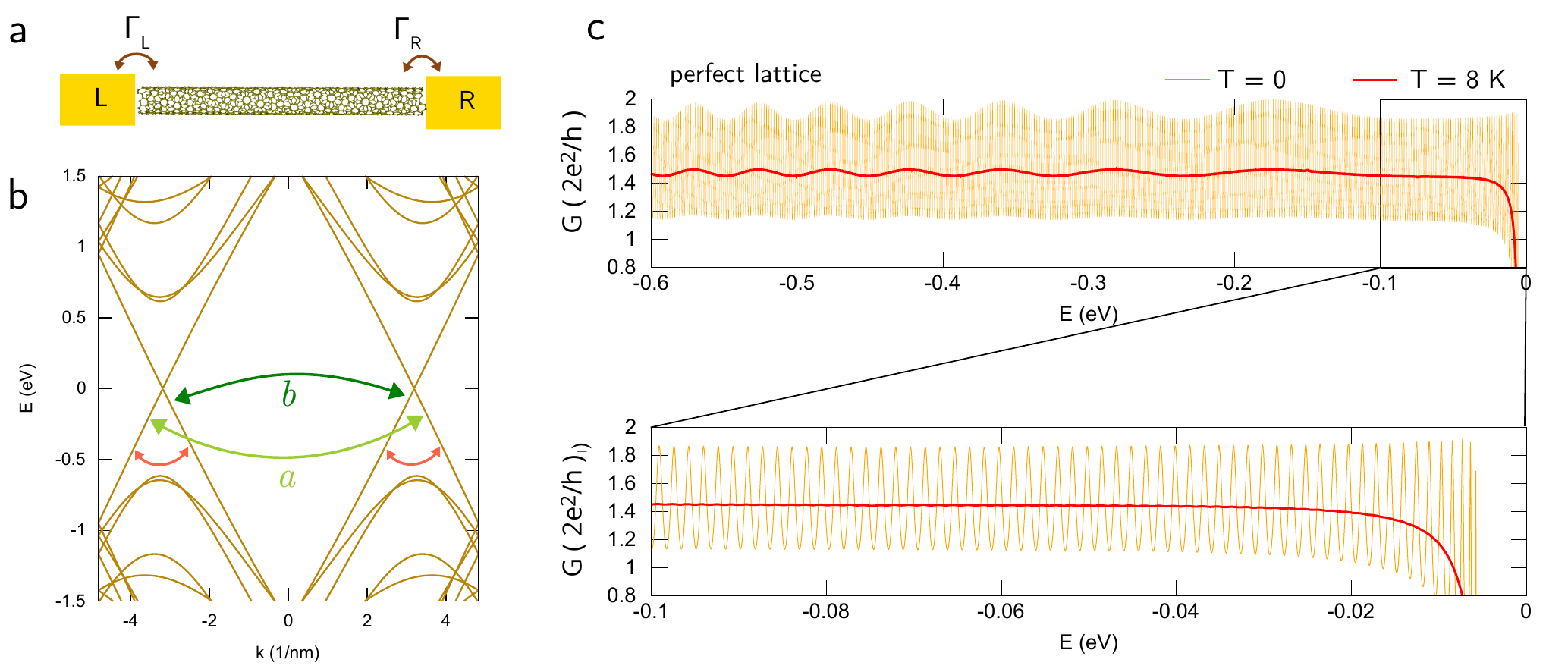}
 \caption{\label{suppfig:FP}
 	Single-particle interference.
(a) Sketch of the calculated setup. The central system with length $L_c=1.04\,\mu$m is contacted to wide band leads by the couplings $\Gamma_L, \Gamma_R$. (b) Low energy dispersion of a (20,5) CNT. The interference channels with higher ($a$) and lower momentum ($b$) are marked by the green arrows. Since this nanotube belongs to the armchair class, the two channels are not independent and can be scattered into each other (this intra-valley scattering is marked by orange arrows). (c) Zero-bias conductance of a (20,5) CNT with the length of 1.04 $\mu$m, comparable to the one in experiment. The orange line is the zero temperature conductance and displays the fast Fabry-P\'{e}rot   oscillations. The red line shows the conductance at $T=$~8K;  no oscillations are discernible close to the band gap (see inset), and only the slow Sagnac oscillation can be seen far from the band edge. }
\end{figure*}

%\begin{figure}[htbp]
%
%
In this section we shortly recall a single-particle approach to Fabry-P\'{e}rot interference and its prediction for a CNT-based electron waveguide. This approach is justified for devices with transparent contacts, when the electron transport through the system is usually too fast to show signatures of charging effects.
Then the conductance assumes overall a high value;  further, low-amplitude periodic oscillations in the conductance  arise from constructive and destructive interference of the electronic trajectories shuttling between the two leads \cite{Liang2001}.
Besides the primary Fabry-P\'{e}rot  interference,  a slow oscillation of the average conductance due to Sagnac interference \cite{Dirnaichner2016,Lotfizadeh2018} arises when the velocities of left- and right-moving electrons do not match in magnitude. \\
In the analytical approach the Fabry-P\'{e}rot interference is described through the different reflection and transmission coefficients of the two modes at the left and right interface, $t_{L/R},r_{R/L}$, respectively. (Since all calculations presented here are at zero bias, instead of $S/D$ from the main text we use the convention of $L/R$ as in Fig.~\ref{suppfig:FP}(a).) In the {\it{absence}} of mixing of the two intervalley channels (orange processes in Fig. \ref{suppfig:FP}(b)) the formula for the overall transmission is given by
\begin{equation}
T(V_g) = \sum_{j=a,b} \frac{2 |t_L|^2|t_R|^2}{1 + |r_L|^2|r_R|^2 - 2|r_L||r_R|\cos(\phi_{j,k}(V_g)) },
\label{eqS1}
\end{equation}
where $j$ labels the two independent channels for interference marked in Fig.~\ref{suppfig:FP}(b) by green arrows, and $\phi_{j,k}(V_g) = (|k_{j,l}(V_g)| + |k_{j,r}(V_g)|) L$ is the phase accumulated by the electron after traversing the nanotube once back and forth, i.e. once on a left-moving branch of the dispersion with momentum $k_{j,l}(V_g)$ and once on the right-moving branch with the dispersion $k_{j,r}(V_g)$. The momentum is related to the gate voltage through the dispersion relation $\varepsilon(k_{j,r/l}) = \alpha eV_g$, where $\alpha$ is the lever arm. The interference pattern in the transmission arises due to the $\cos(\phi_{j,k}(V_g))$ term.\\
Reproducing the experimental transmission curves requires the knowledge of the reflection and transmission coefficients $t_{L/R},r_{L/R}$, yielding four different parameters to adjust. Further, the simple formula \ref{eqS1} cannot account for the beating observed in the experiment  due to combined intravalley and intervalley scattering \cite{Dirnaichner2016}. Hence we turn to a numerical calculation of transmission, using a single particle Green's functions approach,\cite{Datta95} with just the tunnel couplings $\Gamma_L$ and $\Gamma_R$ to the left and right lead, respectively.\\

We chose for the numerical simulation a (20,5) nanotube with the diameter $d = 1.8$~nm and length $L=1.04\, \mu$m, comparable with the experimental parameters. The leads are assumed to be wide band, since the experimental conductance is very high near the band gap.\cite{noteS1} The system is sketched in Fig.~\ref{suppfig:FP}(a). The CNTs band structure in the Dirac regime is shown in Fig.~\ref{suppfig:FP}(b), and the transmission (i.e. the zero temperature linear conductance) in Fig.~\ref{suppfig:FP}(c). It has been obtained with the Landauer-B\"{u}ttiker formula in the Fisher-Lee form,\cite{Datta95}
\begin{equation}
T(E) = \mathrm{Tr}\,[\hat{\Gamma}_L G^R(E)\hat{\Gamma}_RG^A(E)],\hspace{1cm} {\mathrm{with} } \hspace{1cm}
\hat{\Gamma}_{L/R} = \Gamma_{L/R}\mathbbm{1}_c,
\end{equation}
where $\mathbbm{1}_c$ is a diagonal matrix with 1 at the entries corresponding to atoms in contact with the leads and 0 elsewhere. The current is given by
\begin{equation}
I(V_b) = \frac{2e}{h} \int_{-\infty}^\infty d\varepsilon\, \left[f_L(\varepsilon) - f_R(\varepsilon)\right] T(\varepsilon),
\label{current}
\end{equation}
where $f_{L/R}(\varepsilon)=[1+\exp\{(\varepsilon-\mu_{L/R})/(k_BT)\}]^{-1}$ are the Fermi distribution functions of the leads.
%$f(\varepsilon)=[1+\exp(\varepsilon/(k_BT))]^{-1}$.
The lead chemical potentials are given by $\mu_L = \mu_0 + \eta V_b$, $\mu_R = \mu_0 + (\eta-1) V_b$, where $\mu_0=E_F$ is the common Fermi energy of the whole system at zero bias;  $V_b$ is the bias voltage with a possibly asymmetric drop across the nanotube, with the asymmetry encoded in the factor $\eta\in[0,1]$. In the absence of spin-orbit coupling  we assume the two spin channels to be independent and the spin degeneracy is accounted for by the prefactor 2. Eq. (\ref{current}) immediately yields the differential conductance $G_{\rm diff}=dI/dV_b$. The linear conductance follows in the limit of vanishing bias, and it  has the usual form
\begin{equation}
\label{eq:G-temp}
G = \frac{2e^2}{h} \int_{-\infty}^\infty d\varepsilon\, \left.\left(-\frac{\partial f(\varepsilon)}{\partial \varepsilon}\right)\right\vert_{V_b=0} T(\varepsilon).
\end{equation}
We set the zero of the energy at the charge neutrality point of the nanotube. The CNT Fermi energy is then determined by the gate voltage, $E_F = e\alpha V_g$. For $T\approx 0$ the derivative of the Fermi function can be approximated by the Dirac $\delta$ and the linear conductance simplifies even further to
\begin{equation}
G_{T=0} = \frac{2e^2}{h} T(E_F).
\end{equation}
In our setup the linear conductance at $T=0$ is plotted as the orange lines in the Fig.~\ref{suppfig:FP}(c), while the conductance at $T=8$~K (red line) is evaluated through the Eq.~\eqref{eq:G-temp}. The Sagnac interference due to the trigonal warping begins to be visible below the energy of $-0.2$~eV.\\

\begin{figure*}
 \includegraphics[width=0.9\textwidth]{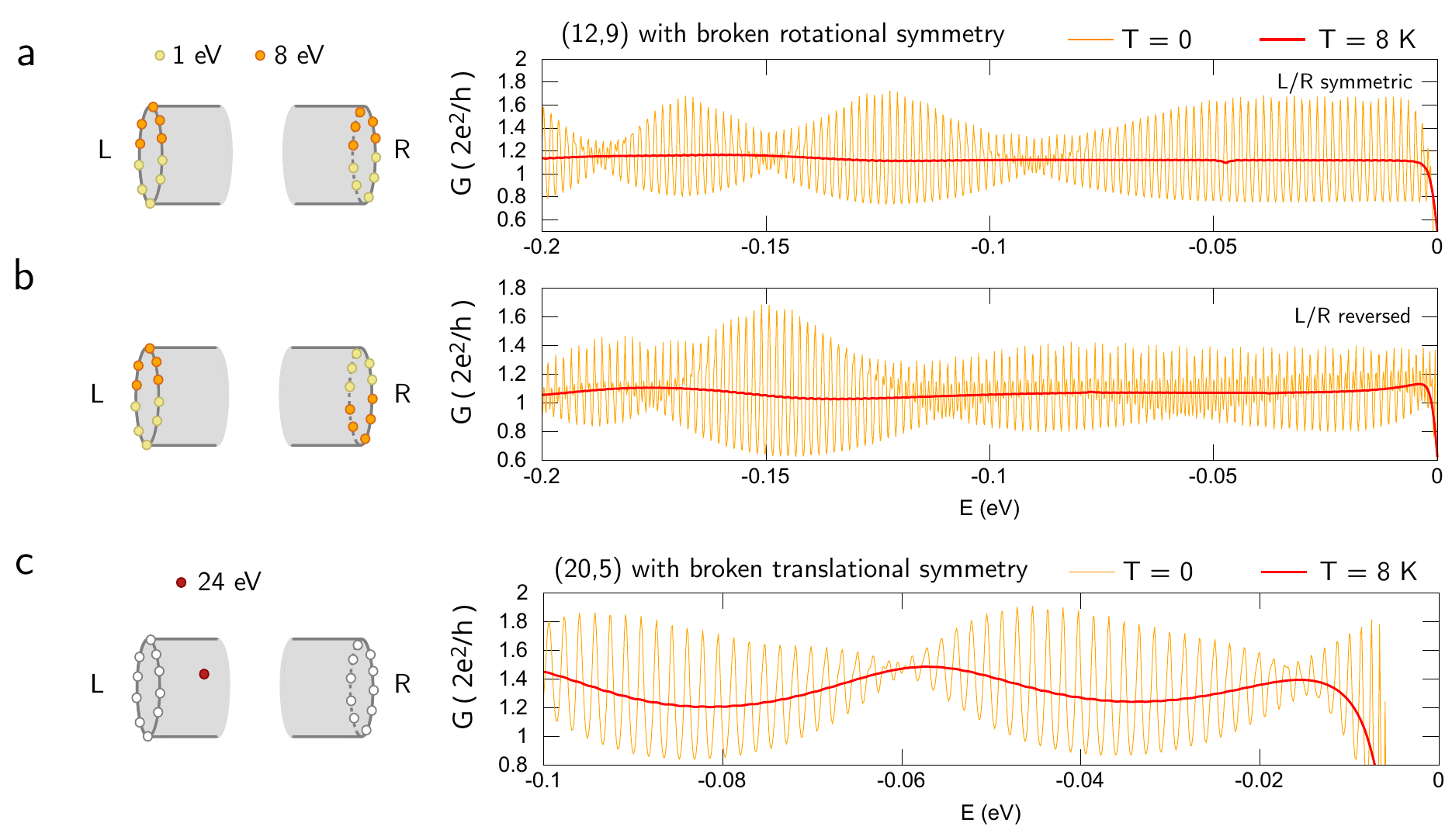}
 \caption{\label{suppfig:FPbroken}
 	Single-particle interference with broken symmetries.
(a),(b) Zero-bias conductance of a (12,9) CNT with length of 1.03~$\mu$m close to the band gap. The uneven tunneling through the top and bottom of the CNT is modelled via additional tunneling barriers at the contact atoms. The two configurations are illustrated schematically, and in both cases the rotational symmetry is broken.
(c) Conductance of a (20,5) CNT with the length of 1.04 $\mu$m near the valence band edge. The lattice contains one adatom at a distance of $\sim36$~nm from the left contact. The adatom is simulated by a local on-site potential of 24~eV.}
\end{figure*}

While the results in Fig.~\ref{suppfig:FP}(c) are obtained for a perfect lattice, the breaking of CNT's symmetries may induce another way to mix the two interference channels. Two such scenarios are illustrated in Fig.~\ref{suppfig:FPbroken}. The rotational symmetry may be broken by different tunneling into the suspended part of the CNT from the top and bottom (in contact with the leads) atoms. In a CNT of the zigzag class this results in mixing the valleys and introducing a modulation of the Fabry-P\'{e}rot interference. This is shown in Fig.~\ref{suppfig:FPbroken}(a),(b) for a (12,9) CNT, with the weaker tunneling at the top of the CNT modelled through increased on-site potential of the contact atoms. In Fig.~\ref{suppfig:FPbroken}(b) the potential configuration at the right lead is reversed with respect to the left lead (physically this would correspond to a CNT which is twisted by half a turn between the left and right lead).\\
The rotational (and translational) symmetry could also be broken by the presence of adatoms in the CNT lattice. The conductance shown in Fig.~\ref{suppfig:FPbroken}(c) has been calculated assuming the presence of an adatom, at the distance of $\sim$36~nm from the left contact, modelled by adding to the Hamiltonian a local on-site energy of 24~eV. The presence of another scattering center and the tiny length scale associated with the adatom-contact distance causes a large scale modulation of the Fabry-Perot interference in the momentum space. \\
In both cases the resulting modification of the Fabry-P\'{e}rot interference reproduces some of the features of the experimental data in Fig. 1 of the main text and in Fig.~\ref{figS1}, hinting that both may be occurring in the experiment.

Because the Fabry-P\'{e}rot interference relies on phase coherence, raising the temperature destroys the oscillation through decoherence, leaving only the slow modulation of the conductance, see Figs.~\ref{suppfig:FP} and \ref{suppfig:FPbroken}. Hence, higher temperature clearly does not introduce the four-time faster oscillations seen in the experiment. This suggests that the low temperature experimental result cannot be simply interpreted in terms of Fabry-P\'{e}rot interference of  {non-interacting} electrons. What we observe in the experiment is rather the interference of quasi-particle excitations of an interacting system.\\

%and with the single particle Green's functions $G^{R/A} = (E\unity - H \mp (\Sigma_R + \Sigma_L))^{-1}$ %calculated numerically in the real space. The self-energies $\Sigma_{L/R}$ are constant and purely imaginary, %related to the $\Gamma_{L/R}$ in the usual way, $\Gamma_{\alpha} = \Sigma^R_{\alpha} - \Sigma^A_{\alpha}$.

\begin{figure*}[htbp]
\begin{center}
\includegraphics[width=0.9\textwidth]{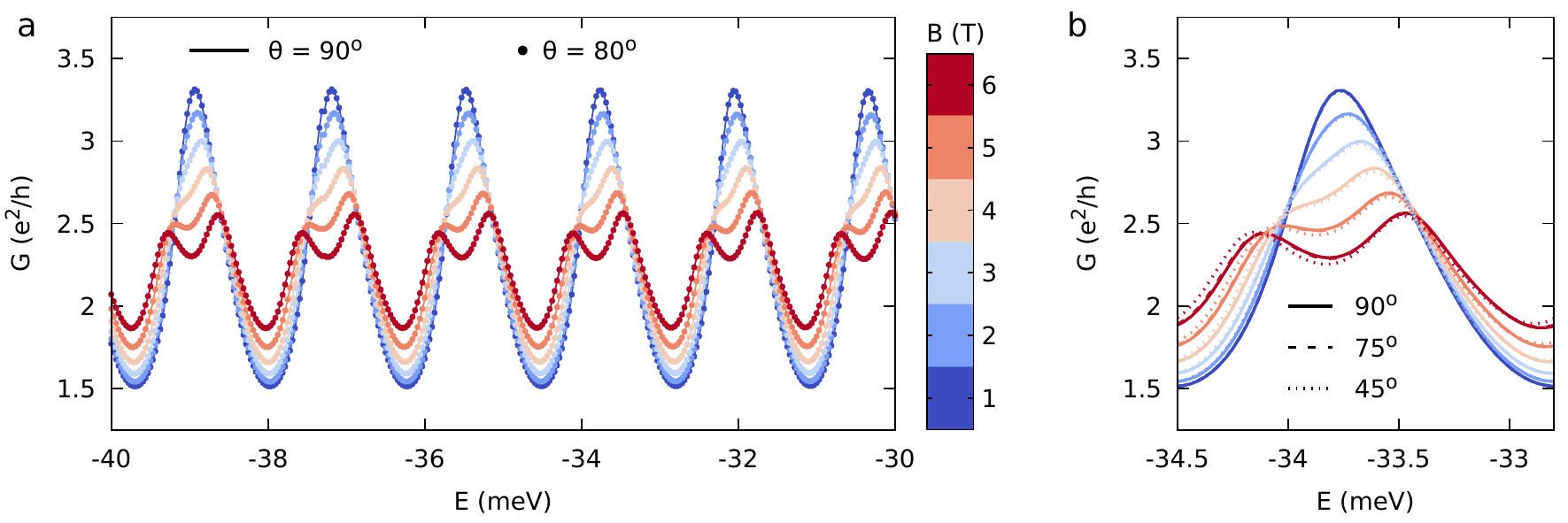}
\end{center}
\caption{\label{suppfig:theoMagnetic}
Fabry-Perot conductance in magnetic field. (a) Linear conductance for a (12,9) CNT with modulated contact potentials for several values of the magnetic field, both perpendicular ($\theta = 90^\circ$) and oblique ($\theta=80^\circ$) to the CNT axis. (The zero field trace is shown in Fig.~\ref{suppfig:FPbroken}a ). The difference between the results for two angle orientations with respect to the CNT axis is indiscernible.
(b) The evolution of one Fabry-Perot peak at different orientations of the magnetic field, showing that unless the field has a significant component parallel to the CNT axis, far from the band gap the orbital response is negligible.}
\end{figure*}

In magnetic field the conductance peaks split, through two possible mechanisms. The field couples to the electron spin via the Zeeman effect and to the valley via the Aharonov-Bohm effect due to a field component parallel to the CNT axis. In perpendicular field we expect only the Zeeman splitting to occur, in tilted field the splitting may be enhanced by the orbital (valley) response. The orbital effect is strongest near the band gap and diminishes for higher  and lower energies.~\cite{jespersen:prl2011}\\
We show in Fig.~\ref{suppfig:theoMagnetic} the results of a numerical simulation of the conductance for a (12,9) CNT, with the same length and configuration of contact potentials as shown in Fig.~\ref{suppfig:FPbroken}a, in perpendicular magnetic field and in field misaligned by $10^\circ$. The orbital response cannot be discerned in the results in Fig.~\ref{suppfig:theoMagnetic}a, and as we can see from a closer inspection of one of the peaks in Fig.~\ref{suppfig:theoMagnetic}b, the magnetic field needs to have a significant component aligned with the tube axis in order to produce even a weak effect. Hence we conclude that the splitting of the conductance peaks in the experiment arises only from the removal of the spin, not valley, degeneracy. The splitting of the conductance peak induced by the magnetic field in Fig.~\ref{suppfig:theoMagnetic} is consistent with the measured peak splitting in Fig. 3c of the main text.
%We expect their splitting to be described by $\Delta E = g\mu_B B$. The ratio $\Delta E/(\mu_B B)$ for $B=6$~T is shown in
%Fig.~\ref{suppfig:theoMagnetic}b). In contrast to the experimental results, the effective $g$ factor is always lower than 2. This suggests that the removal of spin degeneracy in the experiment has further reaching consequences than just the Zeeman splitting of single-particle energy levels. The suppression of spin correlations destroys also the Kondo-like fourfold peaks at low temperature.}

\subsection{Transport with interactions: coherent sequential tunneling for the four-fold degenerate Anderson model}
The single-particle spectrum of a finite CNT is organized into subsets of nearly fourfold-degenerate energy levels, with each quadruplet corresponding to one quantized longitudinal mode. Our starting point is thus the Hamiltonian of a 4-fold degenerate Anderson model, corresponding to one such quadruplet. It has the form $H=H_d+H_T+H_R+H_L$,
where $H_{T}= H_{TL}+H_{TR}$ describes the tunneling coupling of the dot (d) to left (L) and right (R)  electrodes.
The latter are described as an ensemble of non-interacting electrons and captured by the terms $H_L$ and $H_R$.
Finally, the dot Hamiltonian  has the form
\begin{equation}
\label{eq:hamiltonian-dot}
H_d = \sum_j \varepsilon_d n_j + U \sum_{j<k} n_j n_k + \sum_j \alpha e V_g n_j =: \bar{\varepsilon}_d\sum_j n_j + U \sum_{j<k} n_j n_k,
\end{equation}
where the indices run over the quantum numbers of each of the four degenerate states. Further, $\varepsilon_d$
is the single-particle energy,  $V_g$  the gate potential, and $\alpha$ is the lever arm of the quantum dot. In a carbon nanotube quantum dot the four-fold degeneracy arises  from the presence of both  valley and spin, but here we will number the degrees of freedom  generally by $j=1,2,3,4$. The Coulomb interaction is denoted by $U$ and it corresponds to the charging energy $E_C$ in the main text. In order to recover the other longitudinal modes of the CNT, we will later extend this Hamiltonian to a sum of such 4-fold degenerate levels, separated by an energy $\Delta E$ which we shall take, following the experiment, to be $\Delta E \simeq U/2$.\\
The energies of the many-body states with $N=0,...4$ electrons are $E(N) = N\bar{\varepsilon}_d + N(N-1)U/2$. The chemical potential for each occupation $N$ is then
\begin{equation}
\mu(N) = E(N) - E(N-1) = \bar{\varepsilon}_d + (N-1)\, U,\hspace{1cm} N = 1,...,4.
\end{equation}
In the following we shall use the equation of motion technique (EOM) originally proposed in Ref.~[\onlinecite{Meir1993}] for the spinful Anderson model to evaluate the retarded single particle Green's functions $\tilde G^R(i,\varepsilon)$. Their knowledge will give us first indications for the current through the  four-fold degenerate interacting Anderson model.
In fact with $\nu (i,\varepsilon) =  -2 \mathrm{Im}\, \tilde{G}^R(i,\varepsilon)$ being the spectral function of level $i$, the  current follows from the Meir and Wingreen formula \cite{Meir1992}
\begin{equation}
\label{current2}
I=\frac{e}{h}\sum_{i=1}^{4} \int_{-\infty}^{\infty}d\varepsilon \frac{\Gamma_{Li}\Gamma_{Ri}}{\Gamma_{Li}+\Gamma_{Ri}}\nu(i,\varepsilon)[f_L(\varepsilon)-f_R(\varepsilon)].
\end{equation}
The coupling asymmetry parameter for the lead $\alpha$ and level $i$ is given by $\gamma_{\alpha i}=\Gamma_{\alpha  i}/\Gamma_i$, with $\Gamma_{i}=\sum_{\alpha =L,R} \Gamma_{\alpha i}$.
The parameter range of interest for the experiment, $U\simeq \Gamma\gg k_BT$,
 is highly non-trivial and in practice not accessible within the truncation schemes proposed in Ref.~[\onlinecite{Meir1993}]. However, the EOM methods enables one to get the exact current  in the non-interacting case; further, it well describes the tunneling dynamics in the coherent tunneling regime  $U\simeq \Gamma \ge k_BT$, as discussed below.
\subsubsection{Atomic limit }
For a 4-fold {\em isolated} system with four single particle states, i.e., $H=H_d$, the equation of motion procedure closes after four iterations, yielding the exact set of coupled equations
\begin{subequations}
\begin{align}
(\varepsilon-\mu(1)+i\eta)\; \tilde{G}^R(i,\varepsilon) & = 1 + U \tilde{D}^R(i,\varepsilon), \\[2mm]
(\varepsilon-\mu(2)+i\eta))\; \tilde{D}^R(i,\varepsilon) & = \sum_{j\neq i}\langle n_j\rangle + U\tilde{F}^R(i,\varepsilon), \\[2mm]
(\varepsilon-\mu(3)+i\eta)\; \tilde{F}^R(i,\varepsilon) & = \sum_{p\neq j,i} \sum_{j\neq i} \langle n_p n_j\rangle
+ U\tilde{H}^R(i,\varepsilon),\\[2mm]
(\varepsilon-\mu(4)+i\eta)\; \tilde{H}^R(i,\varepsilon) & = \sum_{l\neq p,j,i}\sum_{p\neq j,i} \sum_{j\neq i} \langle n_ln_p n_j\rangle,
\end{align}
\end{subequations}
with $\eta=0^+$ a small infinitesimal. The tilded Green's functions in the energy domain  are the Fourier transforms of
the time-dependent Green's functions
\begin{subequations}
	\begin{align}
	G^R(i,t) &= -\frac{i}{\hbar}\theta(t)\langle \{ c_i(t), c_i^\dag\}\rangle,\\[2mm]
	D^R(i,t) &= -\frac{i}{\hbar}\theta(t) \sum_{j\neq i} \langle \{ n_j c_i(t),c_i^\dag\}\rangle,\\[2mm]
	F^R(i,t) & = -\frac{i}{\hbar}\theta(t)\sum_{j\neq i}\sum_{p\neq i,j} \langle\{ n_j n_p c_i(t), c_i^\dag\}\rangle,\\[2mm]
	H^R(i,t) &= -\frac{i}{\hbar}\theta(t)\sum_{j\neq i} \sum_{p\neq j,i} \sum_{m\neq p,j,i} \langle\{ n_mn_pn_jc_i(t),c_i^\dag \}\rangle.
	\end{align}
\end{subequations}
Each of the four Green's functions describes adding an electron to the level $i$ if either the dot is empty ($G^R(i,t)$), or already hosts one ($D^R$), two ($F^R$) or three ($H^R$) particles.
Solving this set of coupled equations yields the single particle Green's function $\tilde{G}^R(i,\varepsilon)$, which can be conveniently expressed in the form
\begin{equation}
\tilde{G}^R(i,\varepsilon)=\sum_{n=1}^4 \frac{a_n (i)}{\varepsilon-\mu(n)+i\eta},
\label{G_atomic}
\end{equation}
with the coefficients $a_n$ obeying the sum rule $\sum_n a_n=1$.
Let us introduce the occupation numbers
\begin{equation}
\begin{split}
\bar{N}_{1\Sigma}:= & \sum_{j\neq i} \langle n_j\rangle, \\[2mm]
\bar{N}_{2\Sigma}:= & \sum_{j\neq i} \sum_{p\neq j,i}\langle n_jn_p\rangle, \\[2mm]
\bar{N}_{3\Sigma}:=  & \sum_{j\neq i} \sum_{p\neq j,i} \sum_{l\neq p,j,i}\langle n_jn_pn_l\rangle.
\end{split}
\end{equation}
Then in terms of such occupations the coefficients $a_n(i)$ are given by
\begin{subequations}
	\begin{align}
	a_1(i) &= 1 - \bar{N}_{1\Sigma}(i) + \frac{\bar{N}_{2\Sigma}(i)}{2} - \frac{\bar{N}_{3\Sigma}(i)}{6},\\[2mm]
	a_2(i) &= \bar{N}_{1\Sigma}(i) - \bar{N}_{2\Sigma}(i) + \frac{\bar{N}_{3\Sigma}(i)}{2},\\[2mm]
	a_3(i) &= \frac{\bar{N}_{2\Sigma}(i) - \bar{N}_{3\Sigma}(i)}{2}, \qquad a_4(i) =  \frac{\bar{N}_{3\Sigma}(i)}{6}.
	%\\[2mm]
	%a_4(i) &=  \frac{\bar{N}_{3\Sigma}(i)}{6}.
	\end{align}
\end{subequations}
In equilibrium it is possible to evaluate the expectation values $\bar{N}_{n\Sigma}(i)$ using the Lehmann representation \cite{Bruus2005}.  One finds
\begin{equation}
\langle n_i\rangle = \int\frac{d\varepsilon}{2\pi} (-2\,\mathrm{Im}\tilde{G}^R(i,\varepsilon)) f(\varepsilon),
\end{equation}
where $f(\varepsilon) = [1 + \exp \{(\varepsilon-\mu_0)/k_BT)\}]^{-1}$. Note that since we are now working with interacting particles, we replaced $E_F$ with the reference chemical potential $\mu_0$.
%\strike{In the atomic limit, cf. Eq. (\ref{G_atomic}), the spectral function
%$\nu (i,\varepsilon) =  -2 \mathrm{Im}\, \tilde{G}^R(i,\varepsilon)$ has a particularly simple form yielding,}
Using the expression of the $\tilde{G}^R(i,\varepsilon)$ from Eq.~\eqref{G_atomic}, we find
%\begin{equation*}
%\strike{\langle n_i\rangle = \int\frac{d\omega}{2\pi} \nu (i,\varepsilon) f_{}(\varepsilon)
%= \sum_{n=1}^4\int\frac{d\varepsilon }{2\pi}  a_n(i) \delta(\varepsilon-\mu(n)) f_{}(\varepsilon)
%= \sum_{n=1}^4 a_n(i) f_{}(\mu(n)).}
%\end{equation*}
\begin{equation}
\label{eq:nbar-def}
\langle n_i\rangle = \int\frac{d\omega}{2\pi} \nu (i,\varepsilon) f_{}(\varepsilon)
= \sum_{n=1}^4  a_n(i) \int d\varepsilon f(\varepsilon) \delta(\varepsilon-\mu(n))
= \sum_{n=1}^4 a_n(i)\; f(\mu(n)).
\end{equation}
%\add{In the atomic limit the expression for $\bar{n}_n$ is particularly simple,  $\bar{n}_n = f(\mu(n))$. Let us however keep the general $\bar{n}_n$ for future purposes.}
Similar relations hold for the higher Green's functions. Introducing the  shorthand notation $f_{}(\mu(n))=:f_n$, we find
\begin{subequations}
	\label{eq:multiple-occupations}
	\begin{align}
	\sum_{j\neq i} \langle n_j n_i\rangle &= \int\frac{d\varepsilon}{2\pi} \, (-2\,\mathrm{Im}\tilde{D}^R(i,\varepsilon)) f_{}(\varepsilon) = a_2(i)\, f_2 + 2a_3(i)\, f_3 + 3a_4(i)\, f_4,\\[2mm]
	\sum_{p\neq j,i}\sum_{j\neq i} \langle n_pn_j n_i\rangle &= \int\frac{d\varepsilon}{2\pi} \, (-2\,\mathrm{Im}\tilde{F}^R(i,\varepsilon)) f_{}(\varepsilon) =  2a_3(i)\, f_3 + 6a_4(i)\, f_4,\\[2mm]
	\sum_{m\neq p,j,i}\sum_{p\neq j,i}\sum_{j\neq i} \langle n_mn_pn_j n_i\rangle &= \int\frac{d\varepsilon}{2\pi} \, (-2\,\mathrm{Im}\tilde{H}^R(i,\varepsilon)) f_{}(\varepsilon) = 6a_4(i)\, f_4.
	\end{align}
\end{subequations}
For a degenerate model the single particle occupation $\bar{N}_1:= \langle n_i\rangle$ is independent of the index $i$.
Likewise for the double and triple occupations
$\bar{N}_2 := \langle n_j n_k \rangle $ and
$ \bar{N}_3 := \langle n_j n_k n_m \rangle $. This leads to the final result
 \begin{subequations}
 	\label{eq:bar-as}
 	\begin{align}
 	{a}_1(V_g) &= 1 - \left[3\bar{N}_1(V_g)-3\bar{N}_2(V_g)+\bar{N}_3(V_g)\right],\\[2mm]
 	{a}_2(V_g) &= 3\bar{N}_1(V_g)-6\bar{N}_2(V_g)+3\bar{N}_3(V_g),\\[2mm]
 	{a}_3(V_g) &= 3(\bar{N}_2(V_g)-\bar{N}_3(V_g)),\\[2mm]
 	{a}_4(V_g) &= \bar{N}_3(V_g)
 	\end{align}
 \end{subequations}
 together with
 \begin{subequations}
 	\label{eq:Bis}
 	\begin{align}
 	%f_{FD}(\mu(n,V_g)) & = \left[ 1 + \exp\left(\frac{\mu(n,V_g)}{k_BT}\right)\right]^{-1} =: f_n,\quad\quad \textnormal{with}\quad \mu(n,V_g) = -eV_g +nU,\\[2mm]
 	\bar{N}_1(V_g) &= f_1 \left\{1 + 3(f_1-f_2) -3\frac{f_2(f_1-2f_2+f_3)}{1+2 f_2-2 f_3-d(V_g)} + \frac{f_2f_3(f_1-3f_2+3f_3-f_4)}{(1+f_3-f_4)(1+2f_2-2f_3-d(V_g))}\right\}^{-1},\\[2mm]
 	\bar{N}_2(V_g) &= \bar{N}_1(V_g) \frac{f_2}{1+2f_2-2f_3-d(V_g)},\\[2mm]
 	\bar{N}_3(V_g) &= \bar{N}_2(V_g) \frac{f_3}{1+f_3-f_4},\\[2mm]
 	d(V_g) &= \frac{f_3 (f_2-2f_3+f_4)}{1+f_3+f_4}.
 	\end{align}
 \end{subequations}

\subsubsection{Coherent sequential tunneling approximation}
When considering the influence of the coupling $H_T$ to external leads, the set of equations for the single particle Green's function does not close anymore. This requires truncation and approximation schemes to properly account for the interplay of interactions and tunneling.  We assume that the quantum numbers are conserved by the tunneling, i.e.,
$H_{T\alpha}=\sum_{i,k}t_{\alpha k,i}c^\dagger_{i}d_{\alpha k,i}+h.c.$, with $\alpha =L,R$. Further, $c^\dagger_{i}$, $d^\dagger_{k\alpha,i}$ create an electron in the dot and lead, respectively. The quantity $t_{\alpha k,i}$ describes the tunneling between the lead state with its continuous degree of freedom $k$ and the quantum number $i$. The dispersion of the states with quantum numbers $k,i$ in the lead $\alpha$ is given by $\varepsilon_{\alpha k,i}$. The most crude approximation, which is exact for a noninteracting Anderson model ($U=0$) as well as in the atomic limit ($\Gamma\to 0^+$), amounts to truncating the hierarchy of equations for the higher order Green's function $D^R$, $F^R$ and $H^R$ by neglecting some level non-conserving terms (spin-flip terms in the simpler spin-degenerate Anderson model)~\cite{Bruus2005}. In this way the coupling to the
 leads enters only through a self-energy $\Sigma^R$, independent of $U$ and $T$, and defined by
\begin{equation}
\Sigma^R(i,\varepsilon) = \sum_{\alpha k} \frac{|t_{\alpha k,i}|^2}{\varepsilon - \varepsilon_{\alpha k,i}}, \hspace{1cm} \alpha = L,R.
\end{equation}
In this  approximation one finds
\begin{subequations}
	\begin{align}
	(\varepsilon-\mu(1)+\Sigma^R(i,\varepsilon))\; \tilde{G}^R(i,\varepsilon) & = 1 + U \tilde{D}^R(i,\varepsilon), \\[2mm]
	(\varepsilon-\mu(2)+\Sigma^R(i,\varepsilon))\; \tilde{D}^R(i,\varepsilon) & = \sum_{j\neq i}\langle n_j\rangle + U\tilde{F}^R(i,\varepsilon), \\[2mm]
	(\varepsilon-\mu(3)+\Sigma^R(i,\varepsilon))\; \tilde{F}^R(i,\varepsilon) & = \sum_{p\neq j,i} \sum_{j\neq i} \langle n_p n_j\rangle
	+ U\tilde{H}^R(i,\varepsilon),\\[2mm]
	(\varepsilon-\mu(4)+\Sigma^R(i,\varepsilon))\; \tilde{H}^R(i,\varepsilon) & = \sum_{l\neq p,j,i}\sum_{p\neq j,i} \sum_{j\neq i} \langle n_ln_p n_j\rangle.
	\end{align}
	\label{eq:Lorentzian}
\end{subequations}
In the wide-band limit one finds $\Sigma^R(i,\varepsilon) = -i(\Gamma_L+\Gamma_R)/2 = -i\Gamma/2$.
Hence, comparing with the results from the atomic limit, we  obtain  within this simple scheme that the leads induce a temperature independent broadening $\Gamma$. The Green's function then read
\begin{equation}
\tilde{G}^R(i,\varepsilon)=\sum_{n=1}^4 \frac{a_n}{\varepsilon-\mu(n)+i\Gamma/2},
\label{eq:G_Lorentzian}
\end{equation}
with the  coefficients $a_n$  defined as in the atomic limit through  Eqs. (\ref{eq:bar-as}). However, due to the Lorentzian broadening of the Green's functions, cf. Eqs. (\ref{eq:Lorentzian}) and  (\ref{eq:G_Lorentzian}), the
functions $f_n$ yielding the coefficients $\bar{N}_n$ in 	Eqs. (\ref{eq:Bis}) should be replaced by $F_n:= F(\mu(n))$, where
\begin{equation}
\label{eq:int-Loentzian}
F(\mu(n))= \int\frac{d\varepsilon }{2\pi}f(\varepsilon) (-2)\,\mathrm{Im}\left(\frac{1}{\varepsilon-\mu(n)+i\Gamma/2}\right)
= \frac{1}{2} -\frac{1}{\pi}{\rm Im}\Psi\left(\frac{1}{2}+i\frac{\mu(n)-i\Gamma/2-\mu_0}{2\pi k_{\rm B}T}\right),
\end{equation}
where $\Psi(x)$ is the digamma function.

\begin{figure*}
 \includegraphics[width=0.9\textwidth]{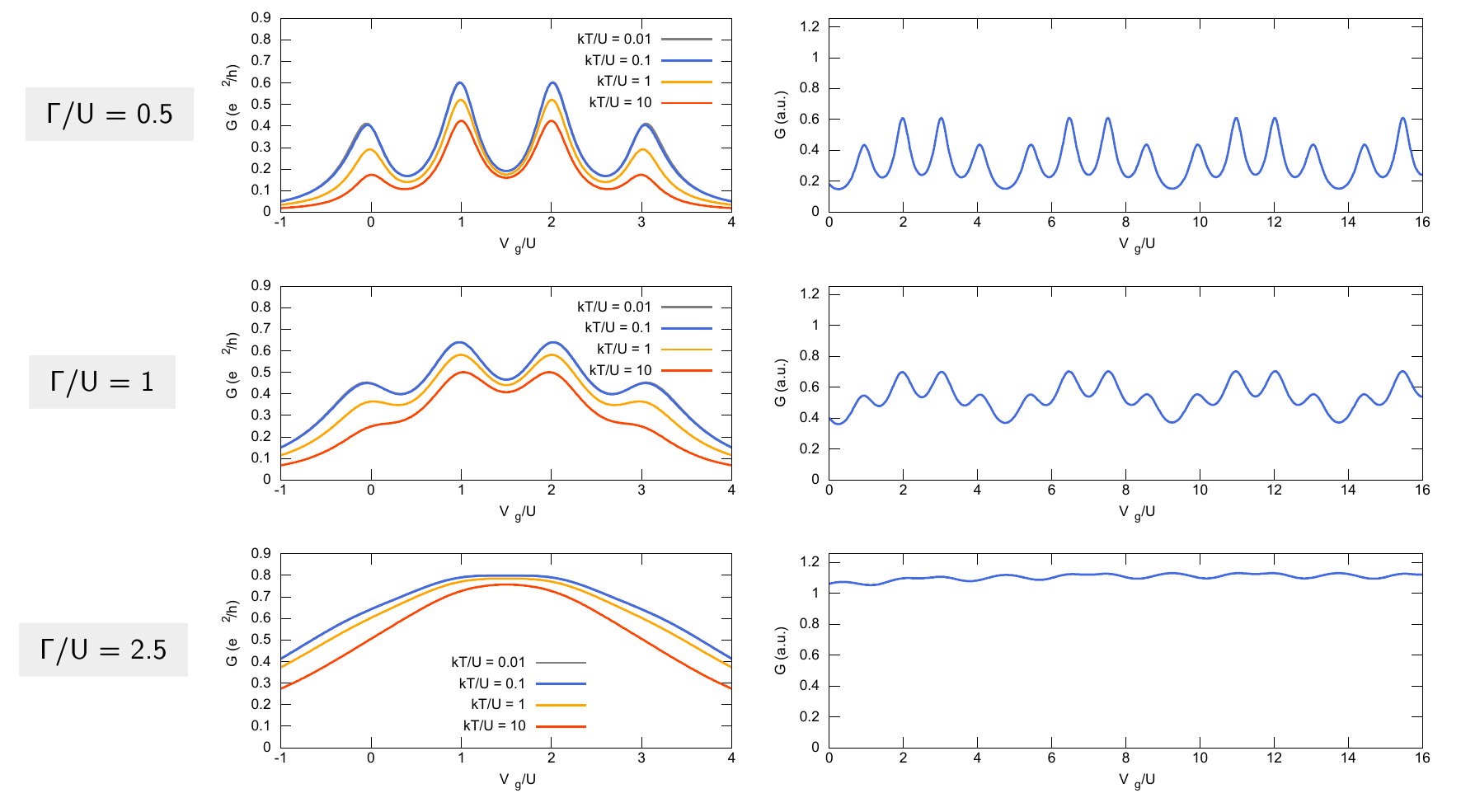}
\caption{\label{suppfig:1stT}Transport through a multilevel Anderson model in the coherent sequential tunneling approximation.  Left column: transport through an Anderson quantum dot with a 4-fold (spin and valley) degenerate single-particle energy level. With increasing broadening $\Gamma$ (approaching the non-interacting limit for $\Gamma/U=2.5$) the four peaks merge into one, but temperature affects the conductance only quantitatively. Right column: conductance through a series of 4-fold degenerate shells with inter-shell spacing  $\Delta E = 0.5U$ and $k_BT/U = 0.1$.
%The central pattern is reminiscent of the experimental zero-bias trace at $T = 4$~K shown in the inset of Fig. \ref{figS3}c;
In the central row the neighboring shells are enhancing the conductance maxima, but the structure of two higher and two lower peaks remains visible. In other words, an enhancement of the central valley similar to what is seen in the experiment is not captured by the coherent approximation.}
\end{figure*}

 The conductance within this Lorentzian scheme is shown in Fig. \ref{suppfig:1stT} for various values of the ratio $\Gamma/U$ and varying temperatures.
Similar to the single-particle interference discussed in the previous section, also in this case the conductance is only moderately dependent on temperature. In particular, a stronger increase of the conductance in the central valley by decreasing temperature, similar to the experimental observations, is not seen (the curves for $k_BT/U = 0.01$ and $k_BT/U=0.1$ are essentially identical). This feature is well known from the studies of the spinful Anderson model within the EOM approach. A temperature dependent self-energy requires accounting for some of the neglected spin-flip contributions \cite{Meir1993,Lavagna2015}. However, an extension which recovers the unitary Kondo limit reached at low temperatures is already very intricate for the spinful case \cite{Lavagna2015}, and becomes intractable  for the four-fold degenerate Anderson model. This generalisation is beyond the scope of this work.

%

%\bibliographystyle{apsrev}
%\bibliography{biblPRLmain}

\begin{thebibliography}{41}
\bibitem[{not()}]{noteS1}
\bibinfo{note}{In a corresponding calculation where the leads were taken to be
  semi-infinite CNTs, the small lead DOS close to the gap always resulted in
  very low conductance minima. The average conductance was rising only at high
  energies.}
\end{thebibliography}
\end{document}